\documentclass[twocolumn,showpacs,preprintnumbers,amsmath,amssymb,superscriptaddress]{revtex4}
\usepackage{epsfig,bm}
\begin{document}
\title{Momentum distribution and correlation of two-nucleon relative motion 
in $^6$He and $^6$Li}
\author{W. Horiuchi\footnote{Research Fellow 
of the Japan Society for the Promotion of Science}}%
\affiliation{Graduate School of Science and Technology, 
Niigata University, Niigata 950-2181, Japan}
\author{Y. Suzuki}
\affiliation{Department of Physics, and Graduate 
School of Science and Technology, Niigata University, Niigata
950-2181, Japan}

\begin{abstract}
The momentum distribution of relative motion between 
two nucleons gives information on the correlation in nuclei. 
The momentum distribution is calculated for both $^{6}$He and $^6$Li 
which are described in a three-body model of $\alpha$+$N$+$N$. 
The ground state solution for the three-body Hamiltonian is obtained
accurately using correlated basis functions.
The momentum distribution depends on the potential model for the 
$N$-$N$ interaction. With use of a realistic potential, 
the $^6$He momentum distribution exhibits a dip around 2\,fm$^{-1}$ 
characteristic of $S$-wave motion. In contrast to this, 
the $^6$Li momentum distribution is very similar to that of the 
deuteron; no dip appears because it is filled with the $D$-wave 
component arising from the tensor force.
\end{abstract}
\maketitle

\section{INTRODUCTION}

In this article we study the momentum distribution and the correlation 
of relative motion of the halo neutrons in two-neutron halo nuclei. 
The importance of the correlation in nuclei is widely recognized; 
obviously the correlation 
plays a vital role in binding a Borromean three-body system which has 
no pairwise bound states. Recently an experiment 
using the technique of 
intensity interferometry~\cite{marques} has been done in order to extract the 
spatial correlation function of the two neutrons in the halo nuclei such 
as $^6$He, $^{11}$Li and $^{14}$Be. In this experiment the 
momenta of the two neutrons and the core 
nucleus are measured after the dissociation of the halo nucleus. 

Another type of experiment has very recently been performed at 
RIKEN to probe the spatial correlation in $^6$He and $^6$Li,
and the analysis of data is in progress~\cite{suda}. 
This experiment utilizes the one-nucleon exchange process in the 
interaction of these $A$=6 nuclei with a 
proton. 
The basic idea is to utilize the well-established one-nucleon exchange
process which is observed at the backward angle in the proton-deuteron elastic
scattering~\cite{sekiguchi}.
Expecting this mechanism to occur in the $A$=6 nuclei,
the cross section for the reaction
$^6$He$(p,dn)\alpha$ ($^6$Li$(p,dp)\alpha$) has been measured
under the backward scattering kinematics of the proton and two-nucleon system
in $^6$He ($^{6}$Li). Under the assumption of a quasi-elastic approximation,
the cross section is expected to be sensitive to the relative momentum between
the two nucleons~\cite{ent}.
Though the reaction mechanism may not be 
as simple as expected, a theoretical analysis of the momentum 
distribution in $^6$He and $^6$Li should be important as a first step 
to understand the physics involved in the experiment. 

There are a number of calculations for the structure of $^6$He in 
various models. We employ a three-body model of $\alpha+N+N$ 
for $^6$He and $^6$Li. It will be important for our purpose to use 
a realistic potential for the interaction between the valence nucleons 
because the correlation between them is primarily determined by the 
$N$-$N$ potential and the distribution at high momentum should be 
sensitive to the short-ranged repulsion and tensor force. 
We refer to Refs.~\cite{funada,csoto} as the most 
relevant calculations which include the tensor force. However, none of the 
literatures including them has considered the momentum distribution 
between the valence nucleons. Of course there are many 
studies which have investigated the momentum distribution of fragments 
such as the core nucleus or the nucleon~\cite{zhukov}.  

The plan of this paper is as follows. In Sec.~\ref{formulation} we 
present our three-body model and some of the details on 
explicitly correlated basis functions which are used to solve 
the three-body problem. Some formulas needed to compare $^6$He and $^6$Li 
properties are given in this section with emphasis on 
the two-nucleon correlation function and the 
momentum distribution. The latter is best defined through the Wigner 
distribution function~\cite{wigner}. One well-known formulation of 
single-nucleon momentum distribution near 
nuclear surface was performed by H\"{u}fner and Nemes~\cite{hufner} 
using the Wigner function. Results of calculation are presented in 
Sec.~\ref{result} together with some discussions. Those included are 
the spectroscopic properties, the two-nucleon correlation functions 
which reveal ``dineutron'' as well as ``cigar-like'' configurations 
and the comparison of the momentum distributions. We will show that 
the tensor force which works differently between $^6$He and 
$^6$Li plays a key role in the momentum distribution around 
2\,fm$^{-1}$. A conclusion is given in Sec.~\ref{conclusion}. In 
Appendix we give a formula to calculate a density matrix or the Wigner 
function using the correlated basis functions.   

\section{FORMULATION}
\label{formulation}

\subsection{Three-body Hamiltonian}
The wave functions for $^{6}$He and $^{6}$Li are determined
from variational calculations for 
the core($\alpha$ particle)+$N$+$N$ three-body system which is 
specified by the Hamiltonian
\begin{equation}
H=T_{\bm{r}}+T_{\bm{R}}+U_1+U_2+v_{12}.
\label{three-body.H}
\end{equation}
The subscripts of the kinetic energies $T$ stand for the relative
distance vector $\bm{r}$ between the two nucleons, and 
the relative distance vector $\bm{R}$ from the $\alpha$ particle 
to the center of mass of the two nucleons. 
The set of Jacobi coordinates $(\bm{r}, \bm{R})$ is called T-type hereafter. 
The potential $U_i$ is the $N$-$\alpha$
potential and $v_{12}$ is the $N$-$N$ potential. 
The $\alpha$ particle is treated as a structureless particle. 

As the two-nucleon potential $v_{12}$, we use a realistic potential, 
G3RS (Gaussian soft core potential with 3 ranges) case 1 
potential~\cite{tamagaki}, 
which contains central (C), spin-orbit (LS) and tensor (T) terms. In fact, 
the G3RS potential contains $\bm{L}^2$ and quadratic 
$\bm{L}\!\cdot\!\bm{S}$ terms in even $L$ waves. Their contribution is 
small; the deuteron energy with (without) these terms is 
$-$2.17 ($-$2.28) MeV. Thus we ignore these $L$-dependent 
terms in what follows. The $D$ state 
probability of the deuteron is 4.8\,$\%$ in the G3RS potential. 
We use the G3RS potential because its Gaussian radial form makes 
a numerical calculation much faster than, e.g., the AV8$^{\prime}$  
potential~\cite{av8}. 
It is instructive for the study of the $N$-$N$ correlated 
motion to compare results of calculation between the realistic potential model 
and an effective potential model which has a mild 
short-ranged repulsion. As such an effective potential we 
employ the Minnesota potential~\cite{minnesota}, abbreviated 
to MN, which has no tensor 
component. This potential renormalizes the effect of the tensor 
force into the central force and reproduces the binding energy and 
the root mean square (rms) radius of the deuteron. 

As for the $N$-$\alpha$ potential $U_i$,
we adopt a phenomenological potential~\cite{kknn}, abbreviated to 
KKNN potential, which is determined so as to simulate 
the nonlocal potential that derives from a microscopic 
calculation in an $N$+$\alpha$ cluster model. The KKNN potential is 
parity-dependent, contains 
the central and spin-orbit components, and 
reproduces very well the low-energy $N$-$\alpha$ scattering phase shifts 
of $S$ and $P$ waves. 
The potential is slightly too repulsive in $D$ and $F$ waves~\cite{aoyama}. 
The Coulomb potential for $p$-$\alpha$ is taken as 
\begin{align}
U^\text{Coul}(r)=\frac{2e^2}{r}\, 
\text{erf}\left(\sqrt\frac{4}{3b^2}r\right),
\end{align}
with $\frac{1}{2b^2}$=0.257\,fm$^{-2}$. The KKNN potential is deep 
enough to support an $S$-wave bound state; the state is considered 
redundant and must be removed in the three-body calculation because 
no bound states exist for $^5$He and $^5$Li. 

\subsection{Variational solution with correlated Gaussians}

Trial wave functions for the ground states of $^6$He and $^6$Li 
are expressed, in $LS$ coupling scheme,   
as a combination of explicitly correlated Gaussians:
\begin{equation}
\Psi_{JM}=\sum_{i=1}^KC_i\Psi_{JM}(\Lambda_i, A_i, u_i),
\label{wf.total}
\end{equation}
with the basis function 
\begin{align}
&\Psi_{JM}(\Lambda\!=\!(LS), A, u, \bm{x})\notag\\
&=(1-P_{12})\left\{\text{e}^{-\frac{1}{2}\tilde{\bm{x}}A\bm{x}}
\left[\mathcal{Y}_{L}(\tilde{u}\bm{x})\chi_S(1,2)\right]_{JM}
\eta_{TM_T}(1,2)\right\}.
\label{basis.fn1}
\end{align}
Here the permutation $P_{12}$ ensures 
the antisymmetry of the valence nucleons.

The basis function is specified by a set of nonlinear parameters, 
the orbital and spin angular momenta $\Lambda$=$(LS)$, a 2$\times$2 
positive-definite, symmetric matrix $A$, and a 2$\times$1 matrix $u$. 
The symbol $\,\tilde{}\,$ indicates the 
transpose of a matrix, and the square bracket $[\mathcal{Y}_{L}\chi_S]$ denotes the 
angular momentum coupling.  The short-hand notation $\tilde{\bm{x}}A\bm{x}$ 
with $\tilde{\bm{x}}$=$(\bm{x}_1,\bm{x}_2)$ 
stands for $A_{11}\bm{x}_1^2$+$2A_{12}\bm{x}_1\!\cdot\!\bm{x}_2$+$
A_{22}\bm{x}_2^2$, where 
the coordinates $\bm{x}_1$ and $\bm{x}_2$, called V-type, 
are the distance vectors
of the valence nucleons from the $\alpha$ particle; 
$\bm{x}_1$=$\bm{R}$+$\frac{1}{2}\bm{r}$ and 
$\bm{x}_2$=$\bm{R}\!-\!\frac{1}{2}\bm{r}$. 
The exponential part of the basis function is rotation-invariant.  
The cross term $A_{12}\bm{x}_1\!\cdot\!\bm{x}_2$ describes explicitly 
the two-nucleon correlation, which is vital to obtain a precise 
solution in a relatively small basis dimension~\cite{svm}.
The angular part of the basis function is expressed by the solid
spherical harmonics, ${\cal Y}_{LM_L}(\tilde{u}\bm{x})=
|\tilde{u}\bm{x}|^LY_{LM_L}(\widehat{\tilde{u}\bm{x}})$, specified 
by a global vector 
$\tilde{u}\bm{x}$=$u_1\bm{x}_1$+$u_2\bm{x}_2$. The ratio 
of $u_1$ to $u_2$ characterizes the coordinate which is responsible for 
the rotation of the system~\cite{svm,usukura}. 
(The norm of $u$, $u_1^2$+$u_2^2$, simply affects the normalization 
of the 
basis function but not the rotation itself, and it may be set to unity.) 
The isospin part of the system is expressed by $\eta_{TM_T}$. 
The $p$-$\alpha$ Coulomb 
potential and the neutron-proton mass difference give 
rise to the isospin impurity in $^6$Li, but its effect is rather 
small~\cite{SKCA,arai1}. We ignore the neutron-proton mass difference 
and consider no isospin mixing in the present study. 
The action of $P_{12}$ on the basis function 
is simple; it maintains the functional form~\cite{svm,ps2} as follows: 
\begin{align}
&P_{12}\text{e}^{-\frac{1}{2}\tilde{\bm{x}}A\bm{x}}
\left[\mathcal{Y}_{L}(\tilde{u}\bm{x})\chi_S(1,2)\right]_{JM}
\eta_{TM_T}(1,2) \notag\\
&=(-1)^{S+T}\text{e}^{-\frac{1}{2}\tilde{\bm{x}}{\bar A}\bm{x}}
\left[\mathcal{Y}_{L}(\tilde{\bar u}\bm{x})\chi_S(1,2)\right]_{JM}
\eta_{TM_T}(1,2),
\label{action.P}
\end{align}
where the symbol $\,\bar{\ }\,$ indicates the interchange of the elements 
of the matrix,  
that is, ${\bar A}_{11}$=$A_{22}$, ${\bar A}_{12}$=${\bar A}_{21}$=$A_{12}$, 
${\bar A}_{22}$=$A_{11}$, and ${\bar u}_1$=$u_2$, ${\bar u}_2$=$u_1$. 
 
A conventional choice for describing 
the rotational motion is a partial wave expansion, 
$[{\cal Y}_{\ell_1}(\bm{x}_1){\cal Y}_{\ell_2}(\bm{x}_2)]_{LM_L}$.  
One introduces a set of important partial waves $(\ell_1 \ell_2)$ 
to obtain a converged solution~\cite{hiyama,funada}. 
See also Ref.~\cite{SKCA} for the 
importance of including high partial waves in the Faddeev calculation for 
$^6$Li. Our angular function looks quite different from this and it is much 
simpler than the partial wave expansion. 
No apparent couplings appear, and the use of the global vector 
greatly simplifies the calculation 
of the Hamiltonian matrix elements~\cite{svm,usukura}. The merit of this angular function 
lies not only in its simplicity but also in its performance. 
As was compared in Ref.~\cite{usukura}, the global vector representation 
gives results as accurately as the partial wave expansion. 
Using the decomposition 
\begin{align}
&\mathcal{Y}_{LM_L}(u_1\bm{x}_1\!+\!u_2\bm{x}_2)\notag\\
&=\sum_{\ell=0}^{L} 
\sqrt{\frac{4\pi (2L\!+\!1)!}{(2\ell \!+\!1)!(2L\!-\!2\ell
 \!+\!1)!}}\notag\\
&\quad\times u_1^{\ell}u_2^{L-\ell}
[\mathcal{Y}_{\ell}(\bm{x}_1) 
\mathcal{Y}_{L-\ell}(\bm{x}_2)]_{LM_L},
\label{exp.Y}
\end{align}
the partial waves $(\ell_1 \ell_2)$ contained in the global vector part 
are only $(0L), (1\, L\!-\!1),\ldots,(L0)$. A very important point of our 
basis function is that other  necessary 
partial wave contributions are brought 
about by the $\exp(-A_{12}\bm{x}_1\!\cdot\!\bm{x}_2)$ term. When 
the term is expanded as $\sum_n[ (-A_{12})^n/n!] 
(\bm{x}_1\!\cdot\! \bm{x}_2)^n$, each term in the series produces the 
partial wave components of type, 
$(x_1x_2)^{n-\lambda}[{\cal Y}_{\lambda}(\bm{x}_1){\cal Y}_{\lambda}(\bm{x}_2)]_{00}$ 
with $\lambda$=$n,n\!-\!2,\ldots, 1$ or 0. 
When this  
is combined with $[\mathcal{Y}_{\ell}(\bm{x}_1) 
\mathcal{Y}_{L-\ell}(\bm{x}_2)]_{LM_L}$ coming from 
the global vector part, it is clear that the basis function 
(\ref{basis.fn1}) can practically 
contain important partial wave combinations. 

We note that the basis function of Eq.~(\ref{basis.fn1}) 
has a definite parity $(-1)^L$. As the ground states of 
$^6$He and $^6$Li have a positive parity, this basis 
function cannot be used for $L$=1. We need to 
extend the basis function to 
make it possible to include $L$=1 and a positive parity.  
This is made possible by replacing 
${\cal Y}_{LM_L}(\tilde{u}\bm{x})$ by 
$[{\cal Y}_L(\tilde{u}\bm{x})
{\cal Y}_1(\tilde{u'}\bm{x})]_{LM_L}$~\cite{double.gvr}.  
For the case of two nucleons with $L$=1, 
this replacement results in a new basis function 
\begin{align}
&\Psi_{JM}(\Lambda\!=\!(1S), A, \bm{x})=(1-P_{12})\notag\\
&\times\left\{\text{e}^{-\frac{1}{2}\tilde{\bm{x}}A\bm{x}}
\left[\left[\mathcal{Y}_{1}(\bm{x}_1)\mathcal{Y}_1(\bm{x}_2)\right]_{1}
\chi_S(1,2)\right]_{JM}\eta_{TM_T}(1,2)\right\}.
\label{basis.fn2}
\end{align}
Note that $[\mathcal{Y}_{1}(\bm{x}_1)\mathcal{Y}_1(\bm{x}_2)]_{1}$ 
is equal to the vector product $\bm{x}_1\!\times \!\bm{x}_2$ within a 
constant factor.  The operation of $P_{12}$ can be done as in 
Eq.~(\ref{action.P}), with an extra minus sign coming from this 
vector coupling. 

The set of $\Lambda=(LS)$ included in the present calculation is 
summarized as follows: 
\begin{align}
& {\text {For $^6$He}} \ (J^{\pi}=0^+),\ \ \ (LS)=(00),\ (11).\notag\\
& {\text {For $^6$Li}} \ \  (J^{\pi}=1^+),\ \ \ (LS)=(01),\ (10),\ (11),\ 
(21).
\label{channel}
\end{align}
Here the basis function is given by Eq.~(\ref{basis.fn1}) for even $L$ and by 
Eq.~(\ref{basis.fn2}) for odd $L$, respectively. 
Note that the two independent $L$=1 basis states exist for $^6$Li. 

As noted before, the redundant state of the KKNN potential 
has to be eliminated; this elimination
 is a manifestation of the Pauli principle 
for the motion of the valence nucleons. The requirement is met 
by imposing that the trial wave function has no overlap with
the $0s_{1/2}$ bound state of the KKNN potential:
\begin{equation}
\langle 0s_{\frac{1}{2}mm_3}(i)|\Psi_{JM}\rangle=0 
\quad (i=1,2),
\end{equation} 
where the radial coordinate of the $0s_{1/2}$ function is 
$\bm{x}_i$ and $m_3$ is the third component of the isospin of 
the nucleon $i$. The $0s_{1/2}$ bound-state wave functions for the
neutron and the proton 
are only slightly different 
because of the Coulomb potential, and 
thus we ignore their difference in the present calculation. 
The exclusion of the $0s_{1/2}$ component is practically 
achieved by the orthogonal projection
method~\cite{kukulin}. That is, we add in the three-body 
Hamiltonian (\ref{three-body.H}) a nonlocal, pseudo potential,  
\begin{align}
\gamma \sum_{i=1}^2 \sum_{m=\pm\frac{1}{2}}
|0s_{\frac{1}{2}mm_3}(i)\rangle \langle 0s_{\frac{1}{2}mm_3}(i)|, 
\label{pseudo}
\end{align}
and obtain a stable solution for sufficiently large $\gamma$ 
(typically 10$^6$ MeV or larger than that). 

To search for good basis functions, we use the algorithm called 
the stochastic variational method (SVM)~\cite{svm}. 
The SVM increases the basis dimension one by one
by testing a number of candidates that are chosen randomly.
Since each basis function is specified by the parameters 
$(A_{11},A_{12},A_{22}, u_1)$ or $(A_{11},A_{12},A_{22})$, the candidates 
are actually generated by giving random numbers to the parameters 
chosen from physically important multi-dimensional parameter space. 
In this way we determine about 100-200 basis functions for each $\Lambda$.  
The SVM works efficiently 
to take care of both the short-ranged repulsion of 
the realistic force and the elimination of the 
redundant states. 

\subsection{Transformation of coordinate sets}
\label{trans.coord}

In the Faddeev method for a three-body system, three sets of T-type
Jacobi coordinates are used, and each Faddeev component of the 
total wave function, expressed 
in one of the three sets, is expanded in partial waves.  
To specify the basis function, however, we use 
the V-type coordinate $\bm{x}$, which is different from the T-type 
coordinate. The V-type coordinate is a set of ``single particle''(s.p.)-like 
coordinate,  
and it is chosen to make it easy to implement the symmetry of the nucleons. 
See Eq.~(\ref{action.P}). 
The s.p. coordinate is useful to represent the individual motion of the nucleons 
when the correlation term vanishes, i.e., $A_{12}$ is set to zero.  
For example, the $0^+$ ground state wave function of $^6$He, approximated 
in $p$-shell harmonic-oscillator configurations, will be expressed in terms of 
a combination of  the spin-singlet and spin-triplet states
\begin{align}
 \Phi_{0}&={\mathcal N}_0\text{e}^{-\frac{1}{2}a_0(\bm{x}_1^2+\bm{x}_2^2)}
\notag\\&\quad\times\left[\left[
\mathcal{Y}_{1}(\bm{x}_1)\mathcal{Y}_{1}(\bm{x}_2)\right]_{0}\chi_0(1,2)
\right]_{00}\eta_{11}(1,2),\notag \\
 \Phi_{1}&={\mathcal N}_1\text{e}^{-\frac{1}{2}a_1(\bm{x}_1^2+\bm{x}_2^2)}
\notag\\&\quad\times\left[\left[
\mathcal{Y}_{1}(\bm{x}_1)\mathcal{Y}_{1}(\bm{x}_2)\right]_{1}\chi_1(1,2)
\right]_{00}\eta_{11}(1,2),
\label{p.shell.wf}
\end{align}
where ${\mathcal N}_0$ and ${\mathcal N}_1$ are respective normalization 
constants. Here the shell model is extended to 
allow for different size parameters, $a_0$ and $a_1$, for both the components. 

The T-type coordinate $\tilde{\bm{\rho}}$=$(\bm{r}, \bm{R})$ is also convenient to 
impose the exchange symmetry for the two nucleons as 
$P_{12}$ simply changes $\bm{r} \!\to \!-\bm{r}$ in the orbital function. 
(For a system consisting of more than three particles, the symmetry requirement 
will be performed more easily in the V-type coordinate than other Jacobi 
coordinate sets.) 
Another set of coordinates commonly used is 
$\tilde{\bm{\zeta}}$=$(\bm{\zeta}_1, \bm{\zeta}_2)$, called 
Y-type, which is related to the V-type coordinate by 
$\bm{\zeta}_1$=$\bm{x}_1$ and $\bm{\zeta}_2$=$\bm{x}_2\!-\!\frac{1}{A_c+1}\bm{x}_1$, 
where $A_c$ is the mass number of the core nucleus ($A_c$=4 in the 
present case). Each type of coordinate sets emphasizes its characteristic motion. 
As mentioned above, the V-type is suited for describing the s.p. like 
motion around the 
core nucleus~\cite{cosm}, while the T-type coordinate is suitable for 
describing the motion corresponding to $\alpha$+(2$N$) cluster decomposition. 
The Y-type coordinate plays a role similar to the V-type. In the limit of 
large $A_c$, both the Y- and V-type coordinates coincide. 

It should be noted that the basis functions, (\ref{basis.fn1}) 
and (\ref{basis.fn2}), maintain their functional form under the transformation of 
the coordinates. The transformation from $\bm{x}$ to $\bm{\zeta}$, e.g., 
is done by a 2$\times$2 matrix $T$ as $\bm{x}$=$T\bm{\zeta}$. Then, it is 
easy to see that the basis functions change as follows: 
\begin{align}
&\Psi_{JM}(\Lambda, A, u, \bm{x})=
\Psi_{JM}(\Lambda, \tilde{T}AT, \tilde{T}u, \bm{\zeta}), \notag\\
&\Psi_{JM}(\Lambda\!=\!(1S), A, \bm{x})
={\det} T
\Psi_{JM}(\Lambda\!=\!(1S), \tilde{T}AT, \bm{\zeta}). 
\label{transformation}
\end{align}
This flexibility of the basis function enables one to take account of possible 
important correlations of the system,
just by choosing the nonlinear parameters 
suitably in only one particular coordinate set. 

Owing to the transformation property of Eq.~(\ref{transformation}), 
the evaluation of the matrix element of an operator may be made in 
any convenient set of the coordinates.  However, there is one exception.  
The matrix element for an angular momentum-dependent operator 
such as the spin-orbit potential of $U_i$ and the pseudo 
potential~(\ref{pseudo}), if calculated in the V-type coordinate, contains 
some error~\cite{baye}, and it   
should be evaluated in the Y-type Jacobi coordinate set $\bm{\zeta}$. 
If the mass number of the core nucleus $A_c$ is sufficiently large as in our 
previous cases~\cite{horiuchi},  
$\bm{\zeta}_2$ is very well approximated by $\bm{x}_2$, and 
the error becomes small.
In the present case, however, $A_c$ is only four, and we will see 
in Sec.~\ref{spectro} that the error is significant. 

\subsection{Charge and matter radii}

The charge radii of $^6$He and $^6$Li can be calculated by taking into 
account the charge radii of the constituent particles of $\alpha$ 
and $N$ as follows:
\begin{align}
& r_c^2(\text{$^6$He})=\frac{1}{9}\langle \bm{R}^2 \rangle +r_c^2(\alpha)+r_c^2(n), 
\label{he6.radius} 
\\
& r_c^2(\text{$^6$Li})=\frac{2}{9}\langle \bm{R}^2 \rangle + 
\frac{1}{12}\langle \bm{r}^2 \rangle+ \frac{2}{3}r_c^2(\alpha)+\frac{1}{3}r_c^2(n)+
\frac{1}{3}r_c^2(p).
\label{li6.radius}
\end{align}
The charge radii of the constituent particles are 
$\sqrt{r_c^2(\alpha)}$=1.671\,fm~\cite{alphac}, 
$\sqrt{r_c^2(p)}$=0.875\,fm, and 
$r_c^2(n)$=$-$0.1161\,fm$^2$~\cite{pdata}. Equation~(\ref{he6.radius}) 
enables one to deduce model-independent information on the 
$\langle \bm{R}^2 \rangle$ value from 
the experimental charge radius of $^6$He. This will be discussed later.  
We can extend Eq.~(\ref{he6.radius}) to a core+$n$+$n$ system:
\begin{align}
r_c^2(\text{core}\!+\!n\!+\!n)=\left(\frac{2}{A}\right)^2\langle \bm{R}^2 \rangle +
r_c^2(\text{core})+\frac{2}{Z}r_c^2(n),  
\end{align}
where $A$ is the mass number of the system. In 
a $^9$Li+$n$+$n$ three-body model for $^{11}$Li, the recent data on the 
charge radii of $^{9,11}$Li~\cite{li11.radius,Li.rch} 
allow us to deduce $\sqrt{\langle \bm{R}^2 \rangle}$=5.95(3)\,fm. 

It is also interesting to calculate the point matter radius which is defined 
by an rms radius for the point nucleon distribution. 
The point matter radii of $^6$He and $^6$Li are calculated by 
\begin{align}
r_m^2(\text{$^6$He}, \text{$^6$Li})=\frac{2}{9}\langle \bm{R}^2 \rangle + 
\frac{1}{12}\langle \bm{r}^2 \rangle+ \frac{2}{3}r_m^2(\alpha),
\label{matter.radius}
\end{align}
where $r_m^2(\alpha)$ is the mean square matter radius of the $\alpha$ 
particle and its value  can be 
given by $r_m^2(\alpha)$=$r_c^2(\alpha)\!-\!r_c^2(p)\!-\!r_c^2(n)$ 
under the isospin symmetry that the protons and the neutrons in 
the $\alpha$ particle have the same mean square radius. In what follows 
we use $\sqrt{r_m^2(\alpha)}$=1.464\,fm.
Using Eqs.~(\ref{li6.radius}) 
and (\ref{matter.radius}) enables us to relate the matter radius of 
$^6$Li to its charge radius:
\begin{align}
r_m^2(^6\text{Li})=r_c^2(^6\text{Li})-r_c^2(n)-r_c^2(p).
\label{li.matter.radius}
\end{align}
The matter radius of $^6$Li is thus expressed by only measurable quantities.

\subsection{Electric quadrupole moment}

The electric quadrupole moment operator is 
\begin{align}
\hat{Q}=\sqrt\frac{16\pi}{5}\sum_i^{A}e_i\mathcal{Y}_{20}(\bm{r}_i-\bm{X}),
\end{align}
where $e_i$ is the nucleon charge and 
$\bm{X}$ is the center of mass coordinate of the system. 
This operator can be simplified, in the $\alpha$+$n$+$p$ model 
of $^6$Li, to 
\begin{align}
\hat{Q} = \hat{Q}_{\bm{r}}+\hat{Q}_{\bm{R}}
=\sqrt\frac{16\pi}{5}e\left\{\frac{1}{4}\mathcal{Y}_{20}(\bm{r})
+\frac{2}{3}\mathcal{Y}_{20}(\bm{R})\right\}.
\end{align}
Here use is made of the fact that the $\alpha$ particle has spin zero 
and the two valence nucleons have good isospin. The operator 
$\hat{Q}_{\bm{r}}$ is the same as the quadrupole moment operator  
for the deuteron, while the second term $\hat{Q}_{\bm{R}}$ corresponds to 
the quadrupole moment operator for the relative motion between the $\alpha$ 
particle and the center of mass of the $np$ system. 

\subsection{Two-nucleon correlation function}

We define the two-nucleon correlation function by 
\begin{align}
\rho(\bm{x}_1, \bm{x}_2)=\frac{1}{2J+1}\sum_{M}
\langle\Psi_{JM} | \bm{x}_1 \bm{x}_2 \rangle \langle 
\bm{x}_1 \bm{x}_2 | \Psi_{JM}\rangle_{\text{ST}}.
\end{align}
Here $\left<\dots\right>_\text{ST}$ indicates that
the integration is to be performed over the spin and isospin coordinates. 
Because of the average procedure over the $z$ component of 
the total angular momentum, the two-nucleon correlation function 
becomes scalar, that is, it is a function of $x_1, x_2$ and $\theta$, 
the angle between $\bm{x}_1$ and $\bm{x}_2$; 
$\rho(\bm{x}_1, \bm{x}_2)$=$\rho(x_1, x_2, \theta)$.   
The normalization of the total wave function $\Psi_{JM}$ is expressed as 
$\iint \rho(x_1, x_2, \theta) 8\pi^2x_1^2x_2^2\sin\theta dx_1dx_2d\theta$=1.  

\subsection{Momentum distribution}
\label{mom.dist}

The momentum and spatial density distributions of a quantum-mechanical 
system are obtained through the 
Wigner phase-space distribution function~\cite{wigner}. The Wigner 
function is concisely expressed in terms of the density matrix. 
Since we are interested in the momentum distribution for the relative motion between 
the valence nucleons, we introduce the density matrix with respect to 
the relative distance vector $\bm{r}$, one of the T-type coordinates:
\begin{align}
\varrho(\bm{r},\bm{r}^\prime)=\frac{1}{2J+1}\sum_{M}
\int\langle\Psi_{JM} | \bm{r}^\prime \bm{R} \rangle \langle 
\bm{r} \bm{R} | \Psi_{JM}\rangle_{\text{ST}}d\bm{R},
\end{align}
where $\left<\bm{r}\bm{R}|\Psi_{JM}\right>$ is obtained from
$\left<\bm{x}_1\bm{x}_2|\Psi_{JM}\right>$ through
the replacement (\ref{transformation}) 
with $T=\begin{pmatrix}\frac{1}{2} &1\\ -\frac{1}{2}&1\end{pmatrix}$.
We define the density matrix by taking the average over the $z$ component of the 
total angular momentum.  

The Wigner distribution function is defined through the density matrix as 
\begin{align}
W(\bm{r}, \bm{k})
=\frac{1}{(2\pi)^3}\int \varrho\left(\bm{r}+\frac{\bm{s}}{2},\bm{r}-\frac{\bm{s}}{2}\right)
\text{e}^{i\bm{k}\cdot\bm{s}} d\bm{s}.
\end{align}
The density distribution for the $N$-$N$ relative motion is given 
by the diagonal element of the density matrix 
\begin{align}
\rho(\bm{r})=\varrho(\bm{r},\bm{r})=\int W(\bm{r}, \bm{k}) d\bm{k}, 
\end{align}
and the momentum distribution for the $N$-$N$ relative motion is obtained by 
\begin{align} 
\rho(\bm{k})=\int W(\bm{r}, \bm{k}) d\bm{r}.
\end{align}
These distributions are normalized as $\int \rho(\bm{r}) d\bm{r}\!=\!1$, 
and $\int \rho(\bm{k}) d\bm{k}\!=\!1$. 
A formula to calculate the density matrix and the momentum distribution 
is given in Appendix. 

\section{RESULTS AND DISCUSSIONS}
\label{result}

The input parameters we use are $\hbar^2/m_N$=41.47 MeV\,fm$^{2}$, 
$m_{\alpha}$=$4m_N$, and $e^2$=1.440 MeV\,fm, where $m_N$ and $m_{\alpha}$ 
are the masses of the nucleon and the $\alpha$ particle, respectively. 
The $u$ parameter of the MN potential is set to $u$=1 otherwise
mentioned. No spin-orbit component of the MN potential is included.  

\subsection{Spectroscopic properties}
\label{spectro}

In a single channel calculation with a specific $\Lambda$=$(LS)$, 
neither (00) nor (11) channel produces a bound 
state of $^{6}$He below the $\alpha$+$n$+$n$ threshold. 
The case of $^{6}$Li is different, depending on the $N$-$N$ potential. 
When using the MN potential, 
the $\Lambda$=(01) channel gives a bound state, whereas  
in the case of the G3RS potential no single-channel calculation 
produces a bound state below the three-body threshold. This difference
between the two potential models 
is due to 
the tensor force which plays no role in the single (01) channel but 
gains energy through the coupling with different channels. This mechanism is 
similar to binding the neutron and 
the proton in the deuteron with the realistic force. 

\begin{table*}[t]
\caption{Properties of the ground states of $^{6}$He and $^6$Li. 
Energy and length are given in MeV and fm. The $\bm{L}^2$ 
and quadratic $\bm{L}\!\cdot\!\bm{S}$ terms of the G3RS potential 
are neglected. See text for the MMN potential. The Coulomb potential is 
included in the term 
$\left<\right.U_1^{\text{C}}+U_2^{\text{C}}\left.\right>$. 
Experimental data: $E$=$-$0.975, $\sqrt{r_c^2}$=2.054(14)~\cite{hec} 
for $^{6}$He; $E$=$-$3.70, $\sqrt{r_c^2}$=2.55(4)~\cite{lic} or 
2.540(30)~\cite{Li.rch}, $\sqrt{r_m^2}$=2.42(4) for $^6$Li.}
\label{ener}
\begin{center}
\begin{tabular}{l@{\hspace{4mm}}ccc@{\hspace{8mm}}cc@{\hspace{8mm}}ccc}
\hline\hline
&\multicolumn{3}{c}{\hspace{-4mm}$^{6}$He ($0^+$)}
& \multicolumn{2}{c}{\hspace{-4mm}$^6$Li ($1^+$)}&
 \multicolumn{2}{c}{$d$ ($1^+$)}
\\ \cline{2-4}\cline{5-6}\cline{7-8}
& MN & MMN & G3RS &  MN &G3RS &  MN &G3RS\\
\hline
$E$ & $-$0.421 &$-$0.975& $-$0.460&$-$3.91& $-$3.31&$-$2.20&$-$2.28\\
$\left<\right.T_{\bm{r}}\left.\right>$
&10.87&11.87&12.51&17.56&23.28&10.48&16.48\\
$\left<\right.v_{12}^{\text{C}}\left.\right>$
&$-$3.77&$-$4.86&$-$5.62&$-$13.41&$-$7.71&$-$12.69&$-$7.29\\
$\left<\right.v_{12}^{\text{T}}\left.\right>$
&   --   &  --  &0.107&  --   &$-$12.25&--&$-$11.46\\
$\left<\right.v_{12}^{\text{LS}}\left.\right>$
&  --  &  --  &0.021&  --  &  -- &  --  &  --  \\
$\left<\right.T_{\bm{R}}\left.\right>$
&12.47&13.06&12.55&13.29&11.49&--&--\\
$\left<\right.U_1^{\text{C}}+U_2^{\text{C}}\left.\right>$
&$-$17.54&$-$18.51&$-$17.71&$-$19.00&$-$16.44&--&--\\
$\left<\right.U_1^{\text{LS}}+U_2^{\text{LS}}\left.\right>$
&$-$2.46&$-$2.54&$-$2.32&$-$2.34&$-$1.69&--&--\\
\hline
$\sqrt{\left<\bm{r}^2\right>}$&5.05&4.63&4.86&3.48&3.58&3.90&3.96\\
$\sqrt{\left<\bm{R}^2\right>}$&3.89&3.66&3.78&3.49&3.81&--&--\\
$\sqrt{r_m^2}$&2.63&2.49&2.56&2.27&2.39&--&--\\
$\sqrt{r_c^2}$&2.09&2.04&2.07&2.41&2.52&--&--\\
\hline
 $P(00)$&84.7&86.4&87.5&--&--&--&--\\
$P(11)$&15.3&13.6&12.5&1.1&0.8&--&--\\
$P(10)$&--&--&--&6.2&3.9&--&--\\
$P(01)$&--&--&--&91.7&90.3&100&95.2\\
 $P(21)$&--&--&--&1.0&5.0&--&4.8\\
\hline\hline
\end{tabular}
\end{center}
\end{table*}

Full calculations which couple all possible $\Lambda$ channels of 
Eq.~(\ref{channel}) give the results 
listed in Table~\ref{ener}. 
The calculation has been performed in the Y-type coordinates.  
The basis dimension $K$ is 400 for $^6$He and for $^6$Li with the MN 
potential, while it is increased to 600 for $^6$Li with the G3RS potential.
The ground states of $^6$He and $^6$Li obtained with the G3RS potential are 
both underbound by 400-500\,keV, while the MN potential 
underbinds $^6$He 
by more than 500\,keV but overbinds $^6$Li by 200\,keV. 
The underbinding of $^6$He with the MN potential is partly due to 
the fact that the $^1$S$_0$ potential is too repulsive to reproduce 
the experimental $^1$S$_0$ phase shifts. 
The common lack of the binding energies in the case of the G3RS potential 
can be explained by at least three effects: One is 
the deficiency of the attraction in the $D$ and $F$ waves of 
the KKNN potential as discussed 
in Refs.~\cite{aoyama,myo}. According to the latter the energy gain of
$^6$He which is obtained by correcting the potential strength 
is, however, only a few tens of keV. Next is 
the effect of three-body forces~\cite{pieper} though a conclusive 
statement on their magnitude remains open. The third 
effect to be considered is the distortion of the $\alpha$ core to 
$3N+N$ partition or the clustering of the  
$A$=6 nuclei into $^3$H+$^3$H (for $^6$He) and $^3$H+$^3$He (for $^6$Li).  
The coupling of the $\alpha$+$N$+$N$ three-body configuration 
to the distorted configuration produces some energy gain.  
A recent microscopic calculation indicates 
that the two configurations actually have rather large overlap and that 
the energy gain is of order of few hundreds keV in $^6$He~\cite{arai}.  

We here remark on the accuracy of the present calculation by comparing 
to other calculations. For $^6$He 
calculated with the MN potential, our ground state energy, 
$-$0.421\,MeV, is in excellent agreement with the value 
$-$0.42\,MeV obtained by the hyperspherical coordinate method~\cite{theeten}. 
If the calculation is done in the V-type coordinates, (which is not correct 
as mentioned in Sec.~\ref{trans.coord}),  
the ground state energy of $^6$He would go down to $-$0.749\,MeV;
the result is again 
consistent with the values obtained by a Lagrange-mesh 
calculation~\cite{baye} and 
a hybrid T+V model calculation~\cite{aoyama}.  
The corresponding energy 
for the $^6$Li ground state would be $-$4.68\,MeV, instead of $-$3.91\,MeV 
calculated correctly in the Y-type coordinates. These results clearly 
show that the 
correct treatment of the coordinates is important in such light systems 
as $^6$He and $^6$Li.  

Table~\ref{ener} lists the decomposition of the energy into the expectation 
values of the various pieces of the Hamiltonian, the radii and the probability 
$P(LS)$ in \% of finding the $\Lambda$=$(LS)$ component.  
Let $E_{\bm{r}}$ and $E_{\bm{R}}$ denote respectively the energy for 
the relative 
motion of the two nucleons and the energy for the relative motion between 
the $\alpha$ particle and the center of mass of the two nucleons. 
They are defined by 
\begin{align}
& E_{\bm{r}}=\left<\right.T_{\bm{r}}\left.\right> + 
\left<\right.v_{12}^{\text{C}}\left.\right> +
\left<\right.v_{12}^{\text{T}}\left.\right> +
\left<\right.v_{12}^{\text{LS}}\left.\right>, \notag \\
& E_{\bm{R}}=\left<\right.T_{\bm{R}}\left.\right>+
\left<\right.U_1^{\text{C}}+U_2^{\text{C}}\left.\right> +
\left<\right.U_1^{\text{LS}}+U_2^{\text{LS}}\left.\right>. 
\end{align}
In the case of $^6$He, the G3RS and MN potentials give very similar results 
for these energies: $(E_{\bm{r}}, E_{\bm{R}})$=(7.03, $-$7.49)\,MeV for 
G3RS and (7.10, $-$7.52)\,MeV for MN. 
Even each of the expectation values is close to each other as well. 
The tensor and spin-orbit forces between the two neutrons 
play a minor role. 
In contrast to $^6$He, both the potentials exhibit quite different features 
in binding $^6$Li.  The partial energies are 
$(E_{\bm{r}}, E_{\bm{R}})$=(3.32, $-$6.63)\,MeV for G3RS and 
(4.14, $-$8.05)\,MeV for MN, so that the 
difference between the two is modest. However, the content of   $E_{\bm{r}}$ 
in particular is 
quite different between them because of the tensor force and the 
short-ranged repulsion. 
In G3RS the large positive value of $\left<\right.T_{\bm{r}}\left.\right>$ is 
balanced by the tensor contribution 
$\left<\right.v_{12}^{\text{T}}\left.\right>$. 
Though this is similar to the case of 
the deuteron, we see  from the values of 
$\left<\right.T_{\bm{r}}\left.\right>$ and $\sqrt{\left<\bm{r}^2\right>}$ 
that the $np$ pair in $^6$Li is more compressed than the one in the deuteron. 
In spite of these differences the $P(LS)$ values of $^6$Li are 
rather similar in the two potentials except for the $(LS)$=(21) channel
which is largely determined through the tensor coupling to the
dominating channel of $(LS)$=(01). As will be discussed later, the 
$^6$Li quadrupole moment is very sensitive to this coupling. 
The spin-triplet channels occupy about 95\%. 

The $^6$Li charge radius calculated using the G3RS potential 
agrees with the experimental value determined from the electron 
scattering~\cite{lic} and the isotope shift in lithium~\cite{Li.rch}, 
but the MN potential gives 
the charge radius which is small by at least 0.1\,fm. 
This failure of the MN potential 
is due to that the $\sqrt{\left<\bm{R}^2\right>}$ value is predicted 
to be too small. As Eq.~(\ref{li.matter.radius}) shows, 
the use of the $^6$Li charge radius leads to 
an ``experimental'' rms matter radius 
$\sqrt{r_m^2(^6\text{Li})}$=2.42(4)\,fm, which is consistent with that 
determined from the analysis of proton elastic scatterings at intermediate 
energies, $\sqrt{r_m^2(^6\text{Li})}$=2.45(7)\,fm~\cite{alkhazov,egelhof}.  
The theoretical value calculated with the G3RS potential is 2.39\,fm, in 
agreement with experiment. 

Now we turn to the case of $^6$He. 
The charge radius of $^6$He has recently been measured by laser 
spectroscopy technique~\cite{hec}.
Our theoretical value agrees fairly well with the experimental value. 
Using the observed charge radius of $^6$He in Eq.~(\ref{he6.radius}), we 
can deduce the ``experimental'' value of $\sqrt{\left<\bm{R}^2\right>}$, 
which turns out to 
be 3.726(69)\,fm. Our value of 3.78\,fm is very consistent with this value 
considering the fact that the calculated binding energy is small by 500\,keV. 
The rms matter radius of $^6$He extracted 
from the proton elastic scattering is 
$\sqrt{r_m^2(^6\text{He})}$=2.30(7)\,fm~\cite{egelhof}, which is, however, 
smaller than the value (2.48$\pm$0.03\,fm) 
deduced from the interaction cross section analysis~\cite{ozawa}. 
The theory with the G3RS or MN potential 
predicts too large matter radius for $^6$He; this overestimation is 
related to that the calculated binding energy is too small, leading 
to a too large value for $\sqrt{\langle \bm{r}^2 \rangle}$. 
Since the attraction of the MN potential is weak in the 
$^1S_0$ channel, we 
repeated the calculation by increasing the strength of its longest 
range part from $-91.85$ to $-91.85\!\times \!1.07$\, 
MeV so as to reproduce the binding energy. This potential is called 
MMN, and its result is listed in Table~\ref{ener}. 
As expected, the rms matter radius now decreases 
from 2.63 to 2.49\,fm, which is consistent with the values of the 
interaction cross section measurement as well as the fully 
microscopic three-cluster calculations~\cite{arai1,arai}. 

\begin{table}[t]
\caption{Electric quadrupole moments of $^6$Li and the deuteron 
in $e$\,fm$^2$. Values in the parentheses are the contributions 
of the cross terms between the $(LS)\!=\!(01)$ and $(LS)\!=\!(21)$ 
components of the $^6$Li ground state wave function. 
The experimental values are $-0.08178(164)$ $e$\,fm$^2$~\cite{quad6Li} 
for $^6$Li and $0.2860(15)\, e\,$fm$^2$~\cite{quadd} for the deuteron.}
\begin{center}
\begin{tabular}{l@{\hspace{4mm}}cc@{\hspace{8mm}}c}
\hline\hline
&\multicolumn{2}{c}{\hspace{-4mm}$^6$Li (1$^+$)}&$d$ (1$^+$)\\
\cline{2-4}
&MN&G3RS&G3RS\\
\hline
$Q$&$-$0.295& 0.164&0.264\\[-1mm]
 &($-$0.396)& (0.088) & \\
\hline
$Q_{\bm{r}}$&$-$0.034& 0.198&0.264\\[-1mm]
 & ($-$0.094)& (0.160)& \\
$Q_{\bm{R}}$&$-$0.260&$-$0.034&--\\[-1mm]
 & ($-$0.302) & ($-$0.072) &  \\
\hline\hline
\end{tabular}
\end{center}
\label{quad.table}
\end{table}

The electric quadrupole moment of $^6$Li is a long-standing problem. 
The $^6$Li quadrupole moment is negative and small, $-0.08178(164)$ 
$e$fm$^2$~\cite{quad6Li}, while the deuteron quadrupole moment 
is $0.2860(15)$ $e$fm$^2$~\cite{quadd}. Thus it is known that 
the $\alpha\!+\!n\!+\!p$ model which assumes $(0s)^4$ $\alpha$-cluster 
gives the wrong sign even when the tensor force is 
included~\cite{csoto}. See also Ref.~\cite{ryzhikh}. 
Though we do not attempt to solve this enigmatic issue, we just show 
the result of the present model in Table~\ref{quad.table}. As expected, 
the $^6$Li quadrupole moment obtained with the G3RS potential 
turns out to be positive (0.164 $e$fm$^2$). 
Interestingly, the quadrupole moment 
with the MN potential becomes $-0.295$ $e$fm$^2$; this happens because the 
large negative contribution of $Q_{\bm{R}}$ 
is not canceled by the $Q_{\bm{r}}$ value. This contrast between 
the two potential models is 
mainly due to the different contribution of the cross terms between the 
$(LS)\!=\!(01)$ and $(LS)\!=\!(21)$ components of the $^6$Li 
ground state wave function. See the parenthetic values in 
Table II. The mixing of these components is 
due to the tensor force. Therefore the failure of the $^6$Li 
quadrupole moment indicates that we have to consider the effect of the 
tensor force between the core and the valence nucleons. 
It should be noted that a variational 
Monte Carlo calculation gives the quadrupole moment of $-0.23(9)$ 
$e$fm$^2$~\cite{wiringa}.

\subsection{Two-nucleon correlation function}
\label{2n.corr}

\begin{figure*}[t]
\epsfig{file=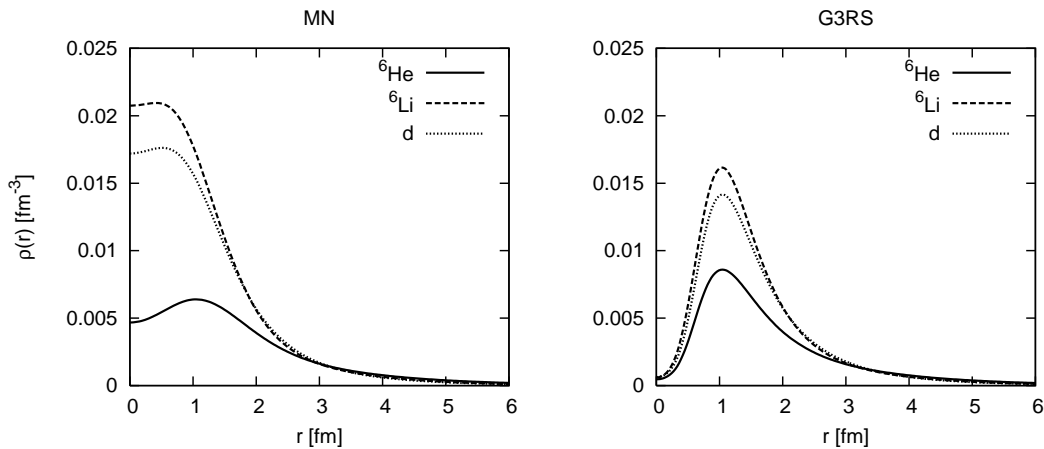,width=14cm,height=5.833cm}
\caption{Density distributions, calculated from the G3RS and MN
 potentials, of the two-nucleon relative motion 
in $^6$He, $^6$Li and the deuteron.}
\label{dens.fig}
\end{figure*}

Figure~\ref{dens.fig} plots the density distributions $\rho({\bm r})$ 
(normalized to unity) of the two-nucleon 
relative motion in $^6$He, $^6$Li and the  deuteron. The densities calculated 
using the G3RS potential (right panel) show central dips due to the 
short-ranged repulsion, but beyond $r$=1.5\,fm they are similar to those 
calculated with the MN potential (left panel). The density of $^6$He 
reaches furthest in the distance, and as a result its density around 
$r$=1$\sim$2\,fm is considerably smaller than that of $^6$Li. 
Comparing the densities between $^6$Li and the deuteron, we see that 
the $np$ relative motion in $^6$Li shrinks compared to that of the 
deuteron (see also the 
$\sqrt{\left<\bm{r}^2\right>}$ value in Table~\ref{ener}). 

\begin{figure}[b]
\epsfig{file=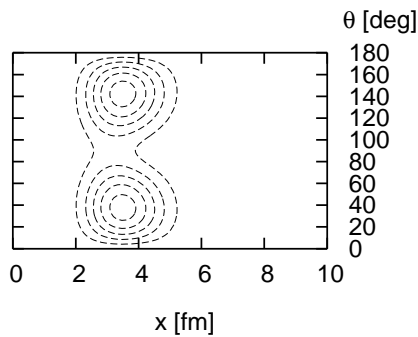,width=6.3cm,height=4.713cm}
\caption{Contour map of the two-nucleon correlation function,
$8\pi^2x^4\sin\theta \rho(x,x,\theta)$, calculated from the 
uncorrelated $p$-shell wave function, Eq.~(\ref{simple.eq}).
The difference between any two neighboring contour levels is 
0.01\,fm$^{-2}$.}
\label{wf2nsimple.fig}
\end{figure}

The correlated motion of the valence nucleons reflects on 
the two-nucleon correlation function $\rho(x_1, x_2, \theta)$. 
Before discussing the correlated motion, we first examine the 
function $\rho(x_1, x_2, \theta)$ which is generated from  
an ``uncorrelated'' basis function $\Phi$ for $^6$He. For this purpose 
we take a combination of the two $p$-shell harmonic-oscillator 
functions (\ref{p.shell.wf}),
\begin{align}
\Phi=\sqrt{1-C^2}\Phi_0+C\Phi_1,
\label{simple.eq}
\end{align}
and determine $(a_0, a_1, C)$ so as to maximize the overlap, 
$|\langle \Phi|\Psi_{00}\rangle|^2$, with the $^6$He ground-state 
wave function 
$\Psi_{00}$ obtained using the G3RS potential. The resulting values are  
$a_0$=0.163\,fm$^{-2}$, $a_1$=0.194\,fm$^{-2}$, and 
$C$=0.402, leading to $|\langle \Phi|\Psi_{00}\rangle|^2$=0.75. 
The simple wave function $\Phi$ has a surprisingly large overlap with 
the realistic wave function $\Psi_{00}$. Though the overlap 
is fairly large, $\Phi$ includes no correlated configurations and indeed 
the energy calculated with $\Phi$ 
is high (8.77\,MeV). The two-nucleon correlation function 
$\rho(x_1, x_2, \theta)$ calculated from $\Phi$ becomes a function of 
$\cos^2 \theta$, so that $\rho(x, x, \theta)$ 
multiplied by $8\pi^2x^4\sin \theta$ is symmetric with respect to 
$\theta$=90$^{\circ}$. See Fig.~\ref{wf2nsimple.fig}. 
An asymmetry with respect to $\theta$=90$^{\circ}$ would indicate the 
presence of correlation in the $A$=6 nuclei. 

\begin{figure*}[t]
\epsfig{file=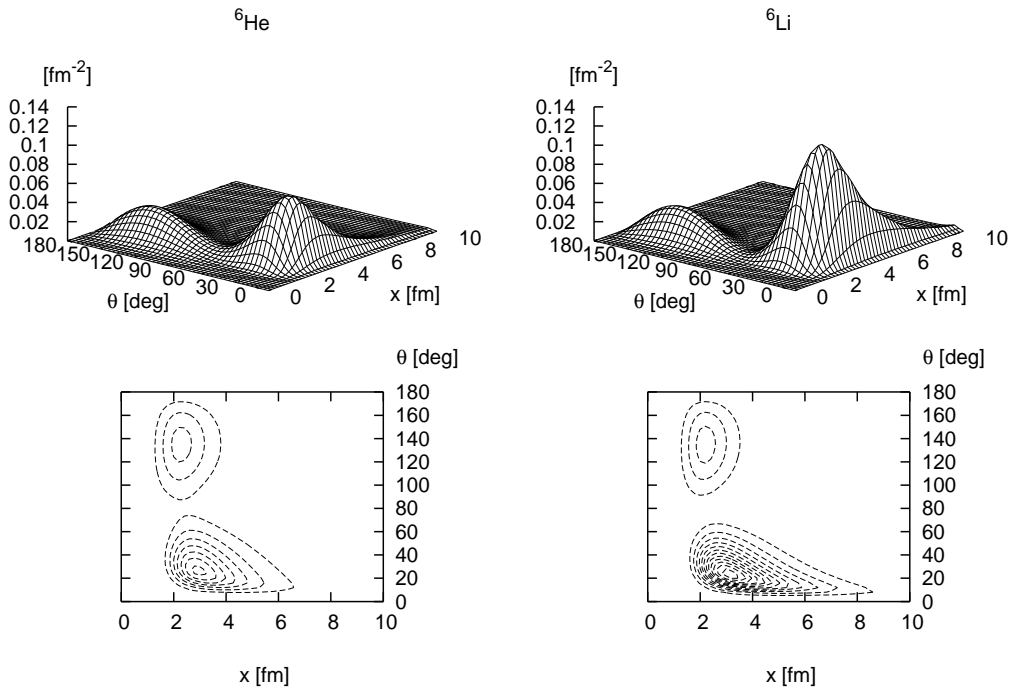,width=14cm,height=10cm}
\caption{Two-nucleon correlation functions, 
$8\pi^2x^4\sin\theta\rho(x,x,\theta)$,
calculated with the G3RS potential for $^6$He and $^{6}$Li.
 The lower panels are their contour
 maps, and the difference between any two neighboring contour levels is
 0.01\,fm$^{-2}$.}
\label{wf2n.fig}
\end{figure*}

Now let us discuss the two-nucleon correlation function which derives from  
the dynamical calculation with the G3RS potential. The MN
potential gives similar two-nucleon correlation function. 
Figure~\ref{wf2n.fig} displays the contour maps of 
$8\pi^2x^4\sin\theta\,\rho(x, x, \theta)$ for $^6$He and $^6$Li. 
In both the cases we clearly see asymmetric patterns with two distinct peaks: 
In $^6$He the highest peak is located  
at $(x, \theta)$=$(2.9, 26^\circ)$ with a height of 0.07\,fm$^{-2}$, 
while in $^6$Li the highest peak is located at $(x, \theta)$=$(3.2, 24^\circ)$ 
with a height of 0.13\,fm$^{-2}$. Here $x$ is given in fm. 
The peak in $^6$Li called the deuteron-like correlation is 
about twice higher than that called the dineutron-like 
correlation in $^6$He. 
Comparing Fig.~\ref{wf2n.fig} with Fig.~\ref{wf2nsimple.fig}, we learn  
that the two-nucleon interaction enhances the asymmetric pattern. 
Another lower peak which shows up at larger angles is called a cigar-like 
correlation. It corresponds to the geometry that the two nucleons 
sit on the opposite sides of the core. This type of peak is located at  
$(x, \theta)$=(2.3, 136$^\circ$) for $^6$He and at 
$(x, \theta)$=(2.4, 136$^\circ$) for $^6$Li, respectively. Their 
heights are both 0.03\,fm$^{-2}$, which is about half of the 
peak height in Fig.~\ref{wf2nsimple.fig}.  

\subsection{Projection to dineutron- and cigar-like configurations} 
\label{projection}

In Sec.~\ref{2n.corr} we discussed the two-nucleon correlated motion 
from the viewpoint of the asymmetric appearance of the dineutron- and 
cigar-like peaks. Here we ask a question of how the 
two peaks appear in relation to the density or rms distance distribution of  
the two nucleons. We expect that the dineutron-like peak in $^6$He 
is formed from those components of the wave function which have smaller 
rms distance, whereas the cigar-like peak is constructed from the rest of 
the components.

\begin{figure}[b]
\epsfig{file=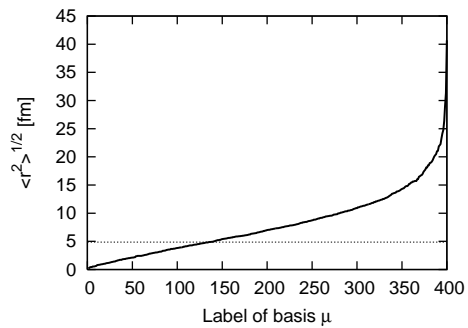,width=6.3cm,height=4.4cm}
\caption{Eigenvalues of the rms distance between the two neutrons,  
$\sqrt{\left<\bm{r}^2\right>}$, calculated from the 
basis functions for $^6$He. The G3RS potential is used. 
Dotted line denotes the 
rms distance (4.86\,fm) of $^6$He.}
\label{eigenrad.fig}
\end{figure}

As Eq.~(\ref{wf.total}) shows, the wave function is given as a combination of 
$K$ independent, nonorthogonal basis functions. By taking a suitable 
linear transformation of these $K$ bases, we can obtain an 
orthonormal set $\Xi_{JM}(\mu)$,
\begin{align}
\Xi_{JM}(\mu)=\sum_{i=1}^KW_{\mu i}\Psi_{JM}(\Lambda_i, A_i,u_i),
\end{align}
with the condition 
$\langle \Xi_{JM}(\mu) | \Xi_{JM}(\mu') \rangle$=$\delta_{\mu, \mu'}$.
It is convenient for the present purpose to choose the coefficients $W_{\mu i}$ in 
such a way that they diagonalize the squared distance $\bm{r}^2$ between 
the two nucleons, that is,  
\begin{align} 
\langle \Xi_{JM}(\mu) | \bm{r}^2 |\Xi_{JM}(\mu') \rangle= 
\langle \bm{r}^2 \rangle_\mu \delta_{\mu, \mu'}.
\end{align}
By arranging the eigenvalues $\left<r^{2}\right>_\mu$
in increasing order ($\mu$=1,2,\ldots, $K$), 
we display in 
Fig.~\ref{eigenrad.fig} the distribution of the rms distance in the case of 
$^6$He, where $K$=400 and the minimum and maximum eigenvalues of 
$\sqrt{\langle \bm{r}^2 \rangle_{\mu}}$ 
are 0.223 and 40.7\,fm, respectively. The fact that the eigenvalues cover 
the wide region from small to large distances indicates that the SVM 
basis selection is efficiently performed to take into account the short-ranged 
correlation as well as the asymptotic behavior. 
We can see that the basis states 
$\Xi_{JM}(\mu)$ of the first 320 members give rather uniform distribution 
of $\sqrt{\langle \bm{r}^2 \rangle_{\mu}}$ up to about 12\,fm, while the 
rest of the basis states cover the eigenvalues of larger rms distances. 

We define two projectors which are orthogonal complements to each other: 
\begin{align}
&P_S=\sum_{\mu=1}^{\kappa}\left|\Xi_{JM}(\mu)\right>\left<\Xi_{JM}(\mu)\right|,
\notag\\
&P_L=\sum_{\mu=\kappa+1}^{K}\left|\Xi_{JM}(\mu)\right>\left<\Xi_{JM}(\mu)\right|,
\label{proj.op}
\end{align}
with $P_S$+$P_L$=1. The projector $P_S$ projects into the subspace spanned 
by those $\Xi_{JM}(\mu)$ which have smaller 
$\langle \bm{r}^2 \rangle_{\mu}$ values, 
while $P_L$ is the projector into the rest which is
spanned by the basis states with larger rms values. 
The total wave function is decomposed into two orthogonal 
components, ``small'' and ``large'', using the projectors 
\begin{align}
\Psi_{JM}=P_S\Psi_{JM}+P_L\Psi_{JM}\equiv\Psi_S+\Psi_L.
\label{proj.eq}
\end{align}
For the sake of simplicity, we choose $\kappa$ in Eq.~(\ref{proj.op}) 
such that 
$\left|\left<\Psi_S|\Psi_{JM}\right>\right|^2$ is as close as to 
0.5. It turns out that 
$\kappa$=88 and the corresponding rms eigenvalue is  
$\sqrt{\left<\bm{r}^2\right>_{\kappa}}$=3.47\,fm.

\begin{figure*}[t]
\epsfig{file=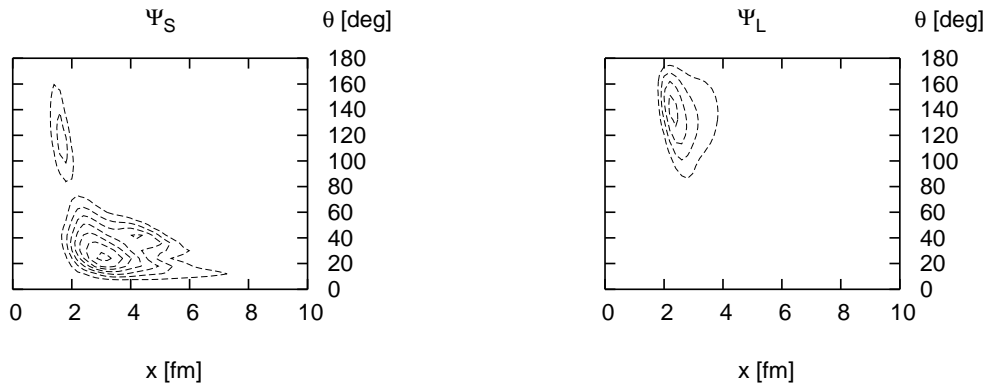,width=14cm,height=5.73125cm}
\caption{Decomposition of the two-nucleon correlation function 
of $^6$He into the small (left panel) and large (right panel) 
components. The G3RS potential is used. 
The difference between any two neighboring contour levels is
 0.01\,fm$^{-2}$.}
\label{corr2nproj.fig}
\end{figure*}

It is now possible to decompose the expectation value of an operator 
$\cal{O}$ into three terms, that is, small, large and their interference terms:
\begin{align}
\left<\Psi_{JM}|\mathcal{O}|\Psi_{JM}\right>&=
\left<\Psi_S|\mathcal{O}|\Psi_S\right>+ \left<\Psi_L|\mathcal{O}|\Psi_L\right>
\notag\\&\quad+\{ \left<\Psi_S|\mathcal{O}|\Psi_L\right>
+\left<\Psi_L|\mathcal{O}|\Psi_S\right>\}.
\label{decomposition}
\end{align}
We apply this decomposition to the two-nucleon correlation 
function to see how the contour map of $^6$He (Fig.~\ref{wf2n.fig}) is 
constructed. 
Plotted in Fig.~\ref{corr2nproj.fig} are those 
contributions to the contour map 
which are calculated with $\Psi_S$ and $\Psi_L$, respectively. 
The contribution of the interference term is found 
to be small and can be safely ignored. Comparing this 
decomposition with the full 
contour map of $^6$He in Fig.~\ref{wf2n.fig}, we can safely conclude that 
the dineutron-like correlation is generated by the small component 
$\Psi_S$ and the cigar-like correlation by the large component $\Psi_S$. 

\subsection{Momentum distribution}

Though the contour map discussed in Secs.~\ref{2n.corr} and 
\ref{projection} shows some correlation effects, it is not clear 
how the correlated features
in $^6$He and $^6$Li are observed 
experimentally. Comparative experiments of the 
intermediate energy proton elastic scatterings on $^6$He and 
$^6$Li~\cite{alkhazov,egelhof} 
have been performed in order to elucidate the matter densities of 
both the nuclei, 
but the analysis of the experimental data is confronted with some 
ambiguities because the scattering is confined to extremely forward 
angles. As mentioned in the Introduction, the measurement of the momentum 
distribution in a special arrangement seems to be accessible 
in the inverse kinematics, providing data which are sensitive 
to the different structures of $^6$He and $^6$Li. 

\begin{figure*}[t]
 \epsfig{file=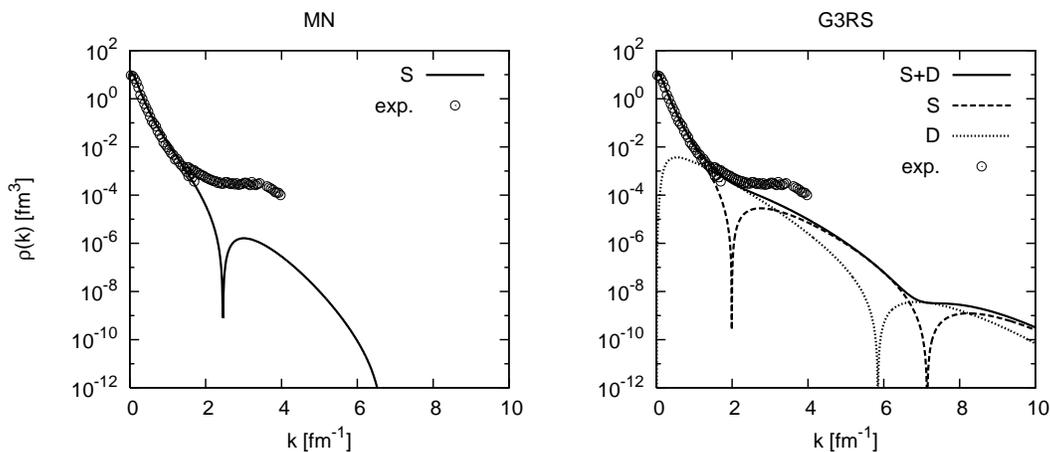,width=14cm,height=5.833cm}
\caption{Momentum distributions of the deuteron calculated with 
the G3RS and MN potentials. Data are taken from Ref.~\cite{deutdist2}.}
\label{dist_d.fig}
\end{figure*}

It is well-known that the momentum distribution of the $np$ relative motion 
in the deuteron shows different behavior in the $S$- and $D$-wave 
contributions. As displayed in the right panel of 
Fig.~\ref{dist_d.fig}, 
the $S$-wave contribution to the momentum distribution 
is peaked at lower momentum and has a dip at $k\!\sim$2\,fm$^{-1}$. 
The $D$-wave component of the deuteron, however, fills the dip. 
This characteristics of the distribution is supported by 
experiment~\cite{deutdist}. In contrast to this,  
the momentum distribution (left panel) obtained with the MN potential 
does show a dip because it has no $D$-wave component, and 
in addition the momentum distribution decreases quickly with increasing $k$ 
because the short-ranged repulsion is not as strong as the G3RS
potential. To compare with experiment at $k$ higher than 2\,fm$^{-1}$, 
however, it is important to include meson exchange currents and 
isobar currents dominated by the $\Delta$ excitation. See 
Ref.~\cite{deutdist2} and Fig.~\ref{dist_d.fig}.

\begin{figure*}[t]
\epsfig{file=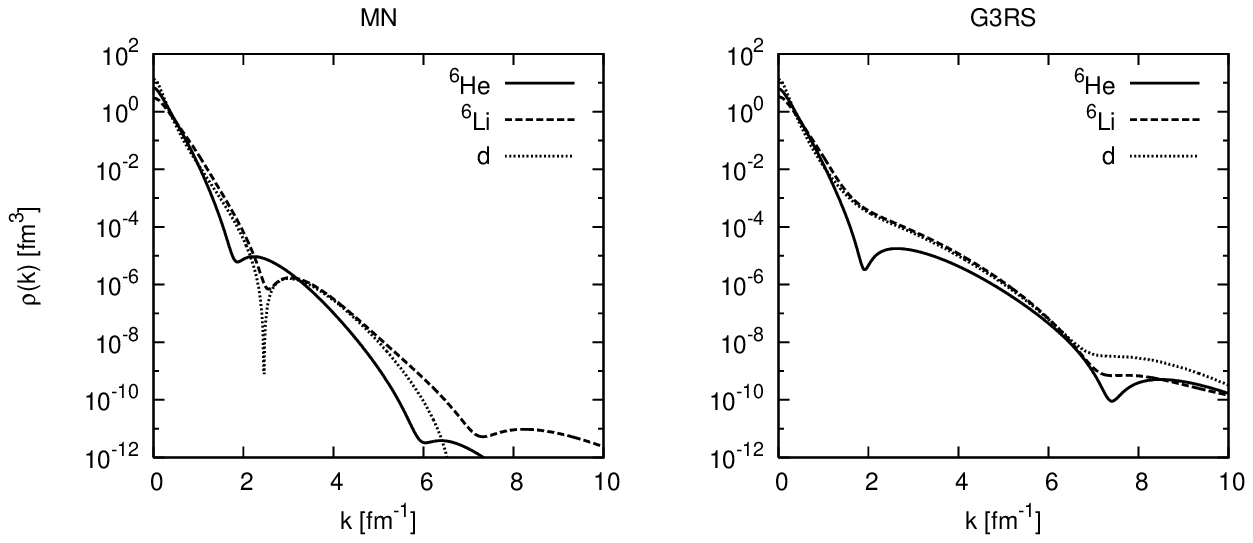,width=14cm,height=5.833cm}
\caption{Momentum distributions, calculated from the G3RS and MN
 potentials, of the valence nucleons 
in $^6$He and $^6$Li. }
\label{dist.fig}
\end{figure*}

The momentum distributions of $^6$He, $^6$Li and the deuteron 
are compared in Fig.~\ref{dist.fig} for the  
G3RS (right panel) and MN (left panel) potentials. 
The realistic potential of G3RS gives the momentum distributions 
characterized as follows: The momentum distribution of $^6$Li is very 
similar to  
that of the deuteron, but the momentum distribution of $^6$He 
differs from them, showing a clear dip at 
$k\!\sim$2\,fm$^{-1}$. These features are understood from the 
difference in the partial wave contents of the $N$-$N$ relative motion; 
$^6$Li contains the $D$-wave component as the deuteron does, whereas 
$^6$He is dominated by the $S$-wave component. 
The most distinctive difference between $^6$He and $^6$Li 
appears around $k\!\sim$2\,fm$^{-1}$.  In this region, however, the 
momentum distribution becomes by four or more order of magnitude 
smaller than that at $k\!\sim\, $0; this may make it hard to 
measure the cross section experimentally. 
If the measurement of the momentum distribution is possible in this region, 
one can learn the role of the tensor force acting between the valence
nucleons, provided that the meson exchange currents and the 
isobar excitations are still not so important.

\begin{figure}[t]
\epsfig{file=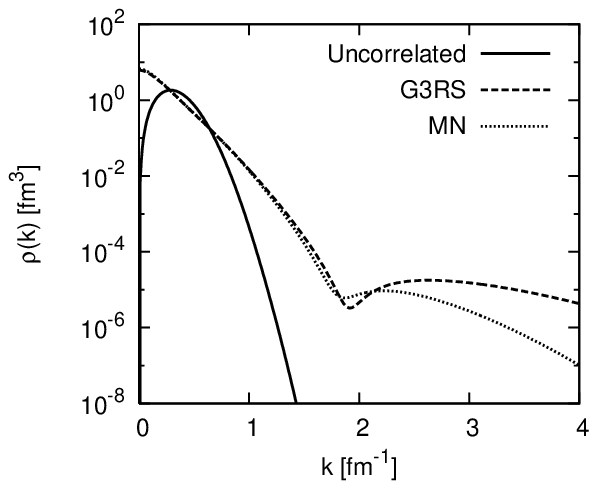,width=6.3cm,height=5.25cm}
\caption{Momentum distributions of the valence nucleons 
in $^6$He for the three different wave functions.}
\label{distsimple.fig}
\end{figure}

Figure~\ref{distsimple.fig} compares the momentum distributions of $^6$He 
corresponding to the three different wave functions, those obtained with 
G3RS and MN, and the uncorrelated one defined in Eq.~(\ref{simple.eq}).  
Both the G3RS and MN distributions are similar up to the dip 
region. Beyond $k\!\sim$2\,fm$^{-1}$ the momentum distribution of 
G3RS surpasses that of MN, which is due to the difference 
in the short-range correlation involved in the two wave functions.  
The uncorrelated wave function gives the 
momentum distribution which is quite different from those of the 
correlated wave functions even at $k\!\sim1\,$fm$^{-1}$. 

\begin{figure*}[t]
\epsfig{file=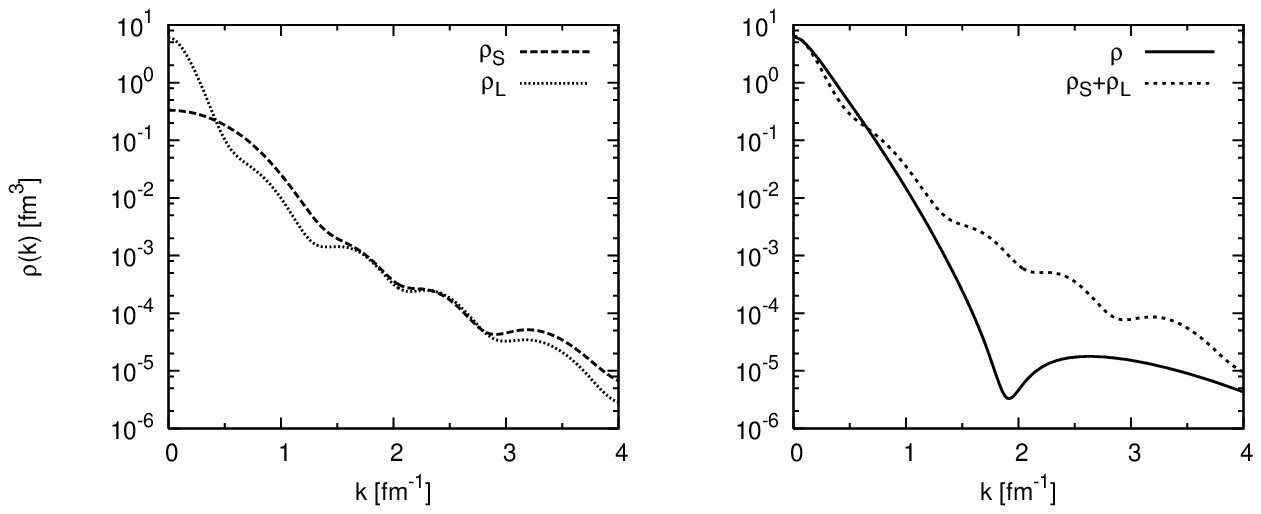,width=14cm,height=5.833cm}
\caption{Momentum distribution of the valence nucleons 
in $^6$He and its decomposition into small and large components. 
See the text for detail. The G3RS potential is used.}
\label{distproj.fig}
\end{figure*}

In Sec.~\ref{projection} we decomposed the ground state wave function 
of $^6$He into $\Psi_S$ and $\Psi_L$, and confirmed that the 
dineutron-like peak is produced by the small component $\Psi_S$, 
while the cigar-like peak by the large component $\Psi_L$. The 
interference term was small. One might 
expect that the momentum distribution may as well be decomposed into 
low and high momentum components. Namely, 
$\Psi_L$ may contribute to the momentum distribution at 
small $k$, while $\Psi_S$ to the high momentum component. 
We examine this expectation in Fig.~\ref{distproj.fig} 
by analyzing the momentum distribution of $^6$He obtained 
with the G3RS potential. 
The left panel shows the partial momentum distributions, $\rho_L$ 
calculated with $\Psi_L$ and $\rho_S$ calculated with $\Psi_S$, respectively.  
The right panel compares the full momentum distribution $\rho$ 
with the incoherent sum of $\rho_L$+$\rho_S$, so that the difference between 
$\rho$ and $\rho_L$+$\rho_S$ is the contribution of the interference terms 
of Eq.~(\ref{decomposition}). We see that the contribution of the 
interference terms can be neglected for $k\!<$1\,fm$^{-1}$.
However, the interference 
contribution becomes important for $k\!>$1\,fm$^{-1}$. In the momentum 
region where the interference can be ignored, the momentum distribution 
is dominated by $\Psi_L$ for $k<0.5$\,fm$^{-1}$ and by 
$\Psi_S$ for $0.5<k<1$\,fm$^{-1}$.

\section{CONCLUSIONS}
\label{conclusion}

To study the correlation and the momentum distribution of the 
two-nucleon relative motion in the ground states of $^6$He 
and $^6$Li, we have described these states in a three-body model of 
$\alpha$+$N$+$N$ where the $\alpha$ particle is assumed to be an 
inert core. We used a parity-dependent $\alpha$-$N$ potential 
which reproduces the low-energy $S$- and $P$-wave phase shifts, and 
two different types of $N$-$N$ interactions as the
potential acting between the two valence nucleons. One is a realistic 
potential which contains the tensor and spin-orbit forces and the other 
is an effective potential which includes no tensor component. 
These were used to compare how much the different $N$-$N$ potentials affect 
the correlation and the momentum distribution.  

We have obtained the solution of the three-body problem by 
approximating the $^6$He and $^6$Li ground state wave functions 
in terms of a combination of explicitly correlated Gaussian basis 
functions. The use of the global vectors to describe a nonspherical orbital 
motion facilitates the calculation of the matrix elements much easier than 
the partial wave expansion, and moreover provides us with 
a solution of high accuracy. 

The energies and rms radii of $^6$He and $^6$Li are compared to 
experiment. The energies calculated with the realistic $N$-$N$ 
potential are 
underbound by 400-500\,keV in both the cases. The charge 
radii of $^6$He and $^6$Li and the matter radius of $^6$Li 
are in fair agreement with the observed values. The $^6$He 
matter radius is predicted to be larger than experiment; 
the result is due 
to the underbinding of the calculated ground state of $^6$He. 
We have analyzed the two-neutron correlated motion in $^6$He 
in order to identify how the dineutron and cigar-like configurations 
are related to the two-neutron relative distance distribution. 
 
The momentum distributions of the $N$-$N$ relative motion have been 
compared between $^{6}$He and $^6$Li. The distributions obtained 
with the effective potential show the pattern characteristic of 
$S$-wave dominance and fall rapidly as the momentum increases. 
In the case of the realistic potential, the momentum 
distribution in $^6$Li is very similar to that of the deuteron. 
That is, both the 
$S$- and $D$-waves contribute to the momentum distribution which  
monotonically decreases with an increasing momentum. In contrast to 
this, the $^6$He momentum distribution is dominated by the $S$-wave, 
showing a clear dip at $k\!\sim\,$2\,fm$^{-1}$. 
The most prominent difference in their momentum distributions thus 
shows up around $k$=2\,fm$^{-1}$. The difference 
between $^{6}$He and $^6$Li is primarily due to whether or not 
the tensor force plays an important role of mixing the $D$-state 
probability between the $N$-$N$ relative motion. We hope that this 
prediction will be tested experimentally.  

\section*{ACKNOWLEDGEMENTS}

The authors thank T. Suda for his interest and valuable discussions. 
They thank D. Baye for useful discussions on the three-body calculations. 
This work was in part supported by Japan-Belgium 
Bilateral Joint Research Project 
of Japan Society for the Promotion of Science and a Grant for 
Promotion of Niigata University Research Projects (2005-2007).

\section*{APPENDIX: CALCULATION OF THE MOMENTUM DISTRIBUTION}

The aim of this appendix is to show a method of calculating the 
momentum distribution for the correlated Gaussians. 
As discussed in Sec.~\ref{mom.dist}, the momentum distribution is 
calculated from the density matrix, so that it is sufficient to show how the 
density matrix is evaluated. We first express the basis functions in 
the T-type coordinates $(\bm{z}_1,\bm{z}_2)$ as discussed in 
Sec.~\ref{trans.coord}, (in this appendix we use 
$(\bm{z}_1,\bm{z}_2)$ instead of $(\bm{r},\bm{R})$ to simplify the notation), 
and write the general form of the orbital part of the correlated Gaussians
as 
\begin{align}
&G_{LM_L}(A,\ell_1,\ell_2,\bm{z}_1,\bm{z}_2)\notag\\
&=\exp{\left(-\frac{1}{2}\tilde{\bm{z}}A\bm{z}\right)}
\left[\mathcal{Y}_{\ell_1}(\bm{z}_1)
\mathcal{Y}_{\ell_2}(\bm{z}_2)\right]_{LM_L}.
\end{align}
The density matrix $\rho$ which we consider here reads as  
\begin{align}
&\rho(\bm{z}_1,\bm{z}_1^{\,\prime})\notag\\
&=\frac{1}{2J+1}\sum_M
\left<\left[G_{L^{\prime}}
(A^\prime,\ell_1^{\,\prime},\ell_2^{\,\prime},\bm{z}_1^{\,\prime},\bm{z}_2)
\chi_{S^\prime}(1,2)\right]_{JM}\right|\notag\\
&\quad\times\left|
\left[G_{L}(A,\ell_1,\ell_2,\bm{z}_1,\bm{z}_2)\chi_S(1,2)\right]_{JM}\right>,
\end{align}
where the integration in the matrix element has to be performed 
for $\bm{z}_2$ as well as the spin coordinates.

Performing the integration over the spin coordinates yields
\begin{align}
&\rho(\bm{z}_1,\bm{z}_1^{\,\prime})=
\delta_{LL^\prime}\delta_{SS^\prime}
\frac{1}{2L+1}\sum_{M_L}\notag\\
&\times\left<G_{LM_L}(A^\prime,\ell_1^{\,\prime},\ell_2^{\,\prime},\bm{z}_1^{\,\prime},
\bm{z}_2)|G_{LM_L}(A,\ell_1,\ell_2,\bm{z}_1,\bm{z}_2)\right>.
\label{orb.int}
\end{align}
Writing the angular part of the right-hand side of Eq.~(\ref{orb.int}) 
explicitly we obtain 
\begin{align}
&\frac{1}{2L+1}\sum_{M_L}
\left[\mathcal{Y}_{\ell_1}(\bm{z}_1)\mathcal{Y}_{\ell_2}(\bm{z}_2)\right]_{LM_L}
\left[\mathcal{Y}_{\ell_1^{\,\prime}}(\bm{z}_1^{\,\prime})
\mathcal{Y}_{\ell_2^{\,\prime}}(\bm{z}_2)\right]_{LM_L}^* \notag\\
&=(-1)^{\ell_1^{\,\prime}+\ell_2^{\,\prime}}\frac{1}{\sqrt{2L+1}}
\sum_\lambda
\begin{bmatrix}
\ell_1&\ell_2&L\\
\ell_1^{\,\prime}&\ell_2^{\,\prime}&L\\
\lambda&\lambda&0
\end{bmatrix}
C(\ell_2 \ell_2^{\,\prime};\lambda)z_2^{2n}\notag\\
&\quad\times\left[
\left[\mathcal{Y}_{\ell_1}(\bm{z}_1)\mathcal{Y}_{\ell_1^{\,\prime}}(\bm{z}_1^{\,\prime})
\right]_\lambda 
\mathcal{Y}_\lambda(\bm{z}_2)
\right]_{00},
\end{align}
where $[\ \ ]$ is the 9$j$ symbol in unitary form and $C$ is the 
coefficient to couple two spherical harmonics with a same argument: 
\begin{align}
C(\ell_2 \ell_2^{\,\prime};\lambda)=\sqrt{\frac{(2\ell_2\!+\!1)
(2\ell_2^{\,\prime}\!+\!1)}{4\pi(2\lambda\!+\!1)}}\langle \ell_2\,0\,
\ell_2^{\,\prime}\,0|\lambda \,0\rangle.
\end{align}
Note that $2n \!\equiv\!\ell_2\!+\!\ell_2^{\,\prime}\! -\!\lambda$ is 
non-negative and even, otherwise the coefficient 
$C(\ell_2 \ell_2^{\,\prime};\lambda)$ vanishes. 
Thus the integration over $\bm{z}_2$ in Eq.~(\ref{orb.int}) is performed as 
\begin{align}
&\int \exp{\left(-\frac{1}{2}az_2^2-\bm{Z}\!\cdot\!\bm{z}_2\right)}
z_2^{2n}\notag\\
&\quad\times\left[
\left[\mathcal{Y}_{\ell_1}(\bm{z}_1)
\mathcal{Y}_{\ell_1^{\,\prime}}(\bm{z}_1^{\,\prime})\right]_\lambda 
\mathcal{Y}_\lambda(\bm{z}_2)\right]_{00}
d\bm{z}_2\notag\\
&=4\pi(-1)^\lambda
\left[\mathcal{Y}_\lambda(\bm{Z})
\left[\mathcal{Y}_{\ell_1}(\bm{z}_1)\mathcal{Y}_{\ell_1^{\,\prime}}(\bm{z}_1^{\,\prime})\right]_\lambda\right]_{00}\notag\\
&\quad\times\sqrt{\frac{\pi}{2}}\frac{(2n)!!}{a^{n+\lambda+\frac{3}{2}}}
L_n^{\left(\lambda+\frac{1}{2}\right)}\left(-\frac{Z^2}{2a}\right)
\exp{\left(\frac{Z^2}{2a}\right)},
\label{app.1}
\end{align}
where $L_n^{\left(\lambda+\frac{1}{2}\right)}$ is the associated
Laguerre polynomial, and 
\begin{align}
a=A_{22}+A_{22}^{\,\prime},\quad
 \bm{Z}=A_{12}\bm{z}_1+A_{12}^{\,\prime}\bm{z}_1^{\,\prime}.
\end{align}

Using the formula (\ref{exp.Y}) the coupling of three $\mathcal{Y}$'s 
in Eq.~(\ref{app.1})  is reduced to
\begin{align}
&\left[\mathcal{Y}_\lambda(\bm{Z})
\left[\mathcal{Y}_{\ell_1}(\bm{z}_1)\mathcal{Y}_{\ell_1^{\,\prime}}(\bm{z}_1^{\,\prime})\right]_\lambda\right]_{00}
\notag\\
&=\sum_k 
\sqrt{\frac{4\pi (2\lambda\!+\!1)!}{(2k \!+\!1)!(2\lambda\!-\!k \!+\!1)!}}
A_{12}^k{A_{12}^{\,\prime}}^{\lambda-k}\notag\\
&\quad\times\sum_\mu
\begin{bmatrix}
k&\lambda\!-\!k&\lambda\\
\ell_1&\ell_1^{\,\prime}&\lambda\\
\mu&\mu&0
\end{bmatrix}
C(k\ell_1;\mu)C(\lambda\!-\!k\, \ell_1^{\,\prime};\mu)\notag\\
&\quad \times z_1^{k+\ell_1}{z_1^{\,\prime}}^{\lambda-k+\ell_1^{\,\prime}}
\left[{Y}_\mu(\bm{z}_1){Y}_\mu(\bm{z}_1^{\,\prime})\right]_{00},
\label{app.2}
\end{align}
with 
\begin{align}
&\left[{Y}_\mu(\bm{z}_1){Y}_\mu(\bm{z}_1^{\,\prime})\right]_{00}
=(-1)^\mu\frac{\sqrt{2\mu\!+\!1}}{4\pi}\notag\\
&\quad\times\sum_{\kappa=0}^{\left[\frac{\mu}{2}\right]}
(-1)^{\kappa}\frac{(2\mu\!-\!2\kappa\!-\!1)!!}{(\mu\!-\!2\kappa)!(2\kappa)!!}
\left(\frac{\bm{z}_1\!\cdot\!\bm{z}_1^{\,\prime}}
{z_1 z_1^{\,\prime}}\right)^{\mu-2\kappa},
\label{app.3}
\end{align}
where $\left[\frac{\mu}{2}\right]$ is the largest integer less than or equal
to $\frac{\mu}{2}$.

Combining Eqs.~(\ref{orb.int})--(\ref{app.3}), 
we obtain the density matrix as a combination of terms 
\begin{align}
z_1^{2p}{z_1^{\,\prime}}^{2p^{\prime}}(\bm{z}_1\!\cdot\!\bm{z}_1^{\,\prime})^q
\exp\left(-\beta z_1^2-\beta^{\prime}{z_1^{\,\prime}}^2-
\gamma \bm{z}_1\!\cdot\!\bm{z}_1^{\,\prime} \right),
\end{align}
where $p,\,p^{\prime}$ and $q$ are all non-negative integers.  

To calculate the momentum distribution we just replace 
$(\bm{z}_1, \bm{z}_1^{\,\prime})$ with $(\bm{r}\!+\!\frac{1}{2}\bm{s}, 
\bm{r}\!-\!\frac{1}{2}\bm{s})$ in the density matrix, multiply 
${\rm e}^{i\bm{k}\cdot\bm{s}}$ and integrate over $\bm{r}$ and $\bm{s}$. 
Renaming $(\bm{r}, \bm{s})$ as $(\bm{z}_1, \bm{z}_2)$ again, 
the integration results in the following form 
\begin{align}
\iint
\text{e}^{-\frac{1}{2}\tilde{\bm{z}}B\bm{z}+i\tilde{\bm{k}}\bm{z}}
z_1^{2n_1}{z_2}^{2n_2}(\bm{z}_1\!\cdot\!\bm{z}_2)^{n_3}d\bm{z}_1d\bm{z}_2, 
\end{align}
which can be performed analytically, where  
$B$ is a $2\!\times\!2$ symmetric matrix, $\tilde{\bm{k}}\bm{z}$
=$\bm{k}\!\cdot\!\bm{z}_1\!-\!\bm{k}\!\cdot\!\bm{z}_2$ and $n_1$, 
$n_2$ and $n_3$ are all non-negative integers.

\end{document}